# *In Situ* Photothermal Response of Single Gold Nanoparticles Through Hyperspectral Imaging Anti-Stokes Thermometry

Mariano Barella,[1] Ianina L. Violi,[1,2] Julian Gargiulo*,[3, †] Luciana P. Martinez,[1] Florian Goschin,[3] Victoria Guglielmotti,[2] Diego Pallarola,[2] Sebastian Schlücker,[4] Mauricio Pilo-Pais,[5] Guillermo P. Acuna,[5] Stefan A. Maier,[3,6] Emiliano Cortes*,[3] Fernando D. Stefani*[1,7]

1. Centro de Investigaciones en Bionanociencias (CIBION), Consejo Nacional de Investigaciones Científicas y Técnicas (CONICET), Godoy Cruz 2390, CABA, Argentina

2. Instituto de Nanosistemas, UNSAM-CONICET, Av. 25 de Mayo 1021, San Martín 1650, Argentina

3. Chair in Hybrid Nanosystems, Nanoinstitute Munich, Faculty of Physics, Ludwig-Maximilians-Universität München, 80799 München, Germany

4. Physical Chemistry I, Department of Chemistry and Center for Nanointegration Duisburg-Essen (CENIDE), University of Duisburg-Essen, Germany

5. Department of Physics, University of Fribourg, Chemin du Musée 3, Fribourg CH-1700, Switzerland

6. The Blackett Laboratory, Department of Physics, Imperial College London, London SW72AZ, United Kingdom

7. Departamento de Física, Facultad de Ciencias Exactas y Naturales, Universidad de Buenos Aires, Int. Güiraldes 2620, CABA, Argentina

*J.Gargiulo@physik.uni-muenchen.de

*Emiliano.Cortes@lmu.de

*fernando.stefani@df.uba.ar

ABSTRACT

Several fields of applications require a reliable characterization of the photothermal response and heat dissipation of nanoscopic systems, which remains a challenging task both for modeling and experimental measurements. Here, we present a new implementation of anti-Stokes thermometry that enables the *in situ* photothermal characterization of individual nanoparticles (NPs) from a single hyperspectral photoluminescence confocal image. The method is label-free, applicable to any NP with detectable anti-Stokes emission, and does not require any prior information about the NP itself or the surrounding media. With it, we first studied the photothermal response of spherical gold NPs of different sizes on glass substrates, immersed in water, and found that heat dissipation is mainly dominated by the water for NPs larger than 50 nm. Then, the role of the substrate was studied by comparing the photothermal response of 80 nm gold NPs on glass with sapphire and graphene, two materials with high thermal conductivity. For a given irradiance level, the NPs reach temperatures 18% lower on sapphire and 24% higher on graphene than on bare glass. The fact that the presence of a highly conductive material such as graphene leads to a poorer thermal dissipation demonstrates that interfacial thermal resistances play a very significant role in nanoscopic systems, and emphasize the need for *in situ* experimental thermometry techniques. The developed method will allow addressing several open questions about the role of temperature in plasmon-assisted applications, especially ones where NPs of arbitrary shapes are present in complex matrixes and environments.

TEXT

Plasmonic nanoparticles (NPs) are widely used in several fields of research due to their outstanding optoelectronic properties.[1,2] Among these properties, they stand out for being remarkably efficient converters of light into heat,[3,4] with absorption cross-sections up to several times their geometrical size. In some cases, plasmon-assisted heating is the primary reason for using plasmonic NPs,[5] such as for photothermal therapy,[6,7] drug delivery and release,[8] solar steam generation,[9] or photothermal microscopy.[10] Alternatively, heating can be an undesired side effect in many other applications like ultra-sensitive (bio)sensing,[11,12] non-linear optics[13] or integrated optoelectronics.[14] In some other areas, the role of thermal effects remains unclear. For example, many efforts have recently been put in the field of plasmon assisted chemistry to disentangle thermal effects from other plasmon derived phenomena such as hot-carriers generation or electromagnetic field enhancement.[15–19] A similar situation occurs in plasmonic optical tweezers, where optical forces are non-trivially intertwined with thermal transport caused by thermophoresis, convection, Brownian motion, or thermoosmosis.[20–23] All these applications of plasmonic NPs have in common the necessity of accurate characterization of the NP photothermal and heat dissipation response.

However, modeling or measuring heat transport in the nanoscale is not straightforward. The challenge in modeling resides in having an accurate description of the irregular interfaces, surface facets or molecular environments[24–27], rendering most thermal simulations only approximate. Measuring the temperature of a nanometric object is also a challenging task.[3] Conventional methods such as infrared lack the necessary spatial resolution. Scanning Thermal Microscopy reaches a 100 nm resolution using a miniature thermocouple at the tip of an atomic force microscope.[28] However, it is technically complex, limited to surfaces, and invasive. Optical methods based on temperature-sensitive luminescent probes are an appealing alternative. Organic dyes, rare-earth doped (nano)crystals, or quantum dots[29] may be used to report temperature through changes in their spectrum,[30] intensity,[31] lifetime or anisotropy[32]. Yet another optical method relies on determining the temperature from variations in the refractive index.[33] However, retrieving accurate temperature measurements of plasmonic NPs using these methods is not always possible. On the one hand, if the probes are not attached to the NPs, the limited spatial resolution leads to temperature values that are a weighted average of the NP and surrounding temperature.[34] On the other hand, if the probes are attached to the NP surface, they may modify both the thermal dissipation and the absorption cross-section of the plasmonic NPs.

Recently, a new optical thermometry method has emerged with the potential to overcome these issues, exploiting the photoluminescence (PL) of plasmonic NPs[35]. In particular, the NP temperature is determined by analyzing the temperature-dependent anti-Stokes (AS) emission in various ways[22,36–39]. The photothermal response of an array of gold nanodisks and inverted pyramids was measured by Xie *et al.*[36] and Hugall *et al.*[37], respectively. Later, Carattino *et al.*[38]

and Cai *et al.*[40] studied individual supported gold nanorods obtained by colloidal synthesis. Jones *et al.*[22] characterized single bowtie gold NPs fabricated on fused silica by electron beam lithography. Finally, Hogan *et. al.*[41] studied Au nanocylinders on a Au film. These implementations of anti-Stokes thermometry differ not only on the target structures but also in the experimental methods and assumptions required for data modeling.

Here, a new implementation of anti-Stokes thermometry that is universally applicable to any NP with detectable PL, and requires no assumption or previous characterization of the NP is presented. The method retrieves the photothermal coefficient of individual NPs from a single PL hyperspectral image. With it, a systematic study of the photothermal response of supported single gold NPs of different sizes in a water environment is performed. The role of the thermal conductivity of the substrate was investigated by performing measurements on NPs supported on glass, sapphire and graphene-coated glass substrates.

RESULTS AND DISCUSSION

**Hyperspectral anti-Stokes nanothermometry**

Figure 1a summarizes the experimental set-up. Heating and PL excitation were performed simultaneously with a continuos wave 532 nm laser in a custom-made sample-scanning confocal microscope. The laser was focused near to its diffraction limit (beam waist $\omega_0 = (342 \pm 5)$ nm). Scanning was achieved with a closed-loop piezoelectric stage. By acquiring a single confocal hyperspectral PL image it is possible to probe the AS emission over a range of temperatures. During scanning, the NP is exposed to a range of irradiance levels (Figure 1b). The integration time at each position was set to 2 s, well beyond the transient time required by the NP to reach thermal steady state, which is in the order of ns to µs. Thus, the measurement at each pixel corresponds to a constant steady-state temperature, with its temperature-dependent PL spectral signature, as depicted in Figures 1c.

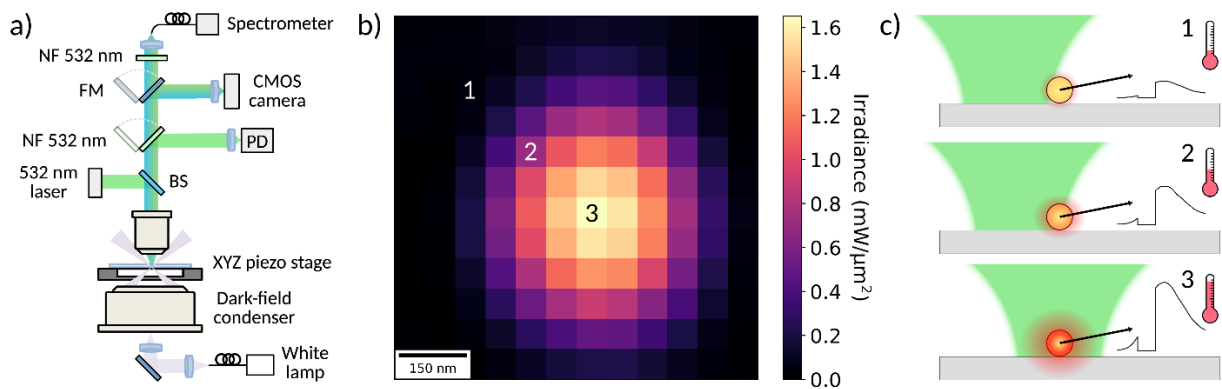

**Figure 1**: Method to obtain spectra vs. irradiance from a single hyperspectral image. (a) Scheme of the experimental set-up. BS: beam-splitter. PD: photodiode. NF: notch filter. FM: flipper mirror. (b) Confocal image of an 80 nm Au NP color coded in terms of irradiance on the NP. (c) Schematic illustration of the heating of a NP during scanning, along with example PL spectra from pixels 1 to 3 marked in (b).

All measurements were performed far from saturation and a linear power dependence of Stokes PL emission was observed, typical of a one-photon process[39] (Figure S1). Exploiting this fact, the irradiance of each pixel was determined as follows: the maximum irradiance of the central pixel was calculated taking into account the incident laser power and the beam geometry. The irradiance of the other pixels was assigned proportionally to their integrated Stokes signal. Then, pixels were grouped according to their irradiance in N equally distributed bins. This improves the signal quality, especially for low irradiance (peripheral) pixels; they have a lower signal but are present in larger numbers. Figure 2a shows a NP PL image color-coded in N=10 irradiance intervals, along with the corresponding PL spectra, background subtracted.

The anti-Stokes (AS) and Stokes emission of each group (i) can be modeled as:

$$I_i^{AS}(\lambda, \lambda_{\text{exc}}, T) \propto I_i^{\text{exc}} f_{PL}(\lambda, \lambda_{\text{exc}}) n(\lambda, \lambda_{\text{exc}}, T) \tag{1a}$$

$$I_i^{S}(\lambda, \lambda_{\text{exc}}) \propto I_i^{\text{exc}} f_{PL}(\lambda, \lambda_{\text{exc}}) \tag{1b}$$

where $I_i^{AS}(\lambda, \lambda_{\text{exc}}, T)$ and $I_i^{S}(\lambda, \lambda_{\text{exc}})$ are the AS and Stokes PL emission at a wavelength $\lambda$, under excitation with a laser of wavelength $\lambda_{\text{exc}}$ and irradiance $I_i^{\text{exc}}$. $f_{PL}(\lambda, \lambda_{\text{exc}})$ is the intrinsic PL emission spectrum, $n(\lambda, \lambda_{\text{exc}}, T)$ is the temperature-dependent distribution of states responsible for the anti-Stokes emission. Extracting the NP temperature, included in $n$, from equation (1a) requires knowledge of $f_{PL}$, which is hard to calculate from first principles because that requires detailed knowledge about the photoluminescence mechanism. Also, several experimental factors have been identified to influence $f_{PL}$ such as excitation wavelength,[42] the electronic band structure of the material,[38] and the photonic density of states (PDOS)[42]. For these reasons, several experimental approaches have been used to determine $f_{PL}$ in the context of AS thermometry. Carattino *et al.* proposed that in the case of Au nanorods $f_{PL}$ follows the surface plasmon resonance with a Lorentzian shape, and used the PL emission measured at a different excitation wavelength to find the parameters of the resonance.[38] Recently, Cai *et al.* approximated $f_{PL}$ with the measured scattering cross section of Au nanorods.[40] However, these approaches require additional characterizations for each NP and make approximations that are not generally valid for every plasmonic NP. For example, these approximations are not valid for nanospheres, where the plasmon resonances are typically broader and closer to interband transitions.[43] As a result, the AS emission spectrum is highly dependent on the excitation wavelength[44] and may not match the scattering spectrum.[40] In order to derive the temperature from $I_i^{AS}$ without needing an explicit expression for $f_{PL}$, a clever strategy was implemented by Xie et al.[36] and followed by Jones et al.[22] and Pensa et al.[45] By taking the ratio $Q_{i,j}^{AS} = \frac{I_i^{AS}}{I_j^{AS}}$ between two spectra obtained at different temperatures, the factor cancels out. This holds if $f_{PL}$ does not depend on temperature, an assumption that is validated by the rather temperature-independent spectra of the Stokes emission (see Figure S2).

The function $n(T)$ expresses the energy distribution of the thermally available states that provide the extra energy for AS emission. It has been already established in the context of AS emission of plasmonic NPs under continuous wave excitation that $n(T)$ is well described by a Bose-Einstein (BE) distribution[36–38,46,47]:

$$n_{BE}(\lambda, \lambda_{\text{exc}}, T) = \left[\exp\left(\frac{E(\lambda) - E(\lambda_{\text{exc}})}{k_B T}\right) - 1\right]^{-1} \quad (2)$$

Where $k_B$ is the Boltzmann constant and $T$ is the temperature (differences between lattice and electronic temperature are negligible under the experimental conditions of this work, see Section 1.4 of the ESI). Furthermore, Hogan et al.[41] used an occupation function including a Bose-Einstein plus a Fermi-Dirac (FD) distribution to estimate the contribution from hot electrons. They found that the maximum contribution of hot electrons is smaller than 3% and that the FD term is only necessary to describe AS emission at energy shifts larger than 0.25 eV (2000 cm$^{-1}$), which is usually beyond the range of detectable anti-Stokes emission of single NPs. Therefore, under the experimental conditions of this work, a BE distribution is a correct representation of $n(T)$. Further discussion on alternative distribution functions can be found in Section 1.5 of the ESI.

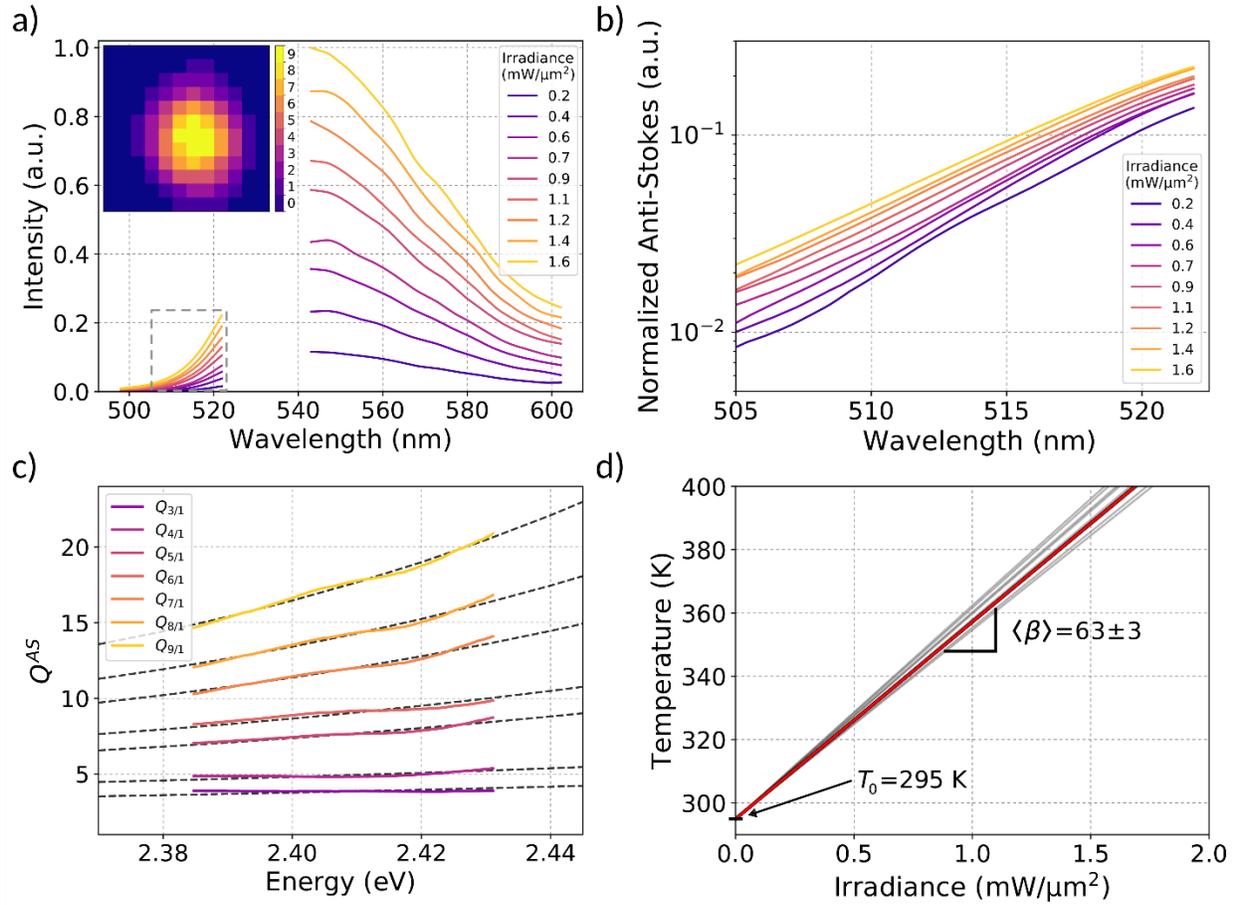

**Figure 2:** Data analysis to obtain the photothermal coefficient $\beta$ from a single hyperspectral PL image. (a) PL spectra of an 80 nm Au NPs at ten different irradiance levels obtained from an hyperspectral confocal image (inset). (b) AS emission of each group normalized by their excitation irradiance. (c) Color solid lines: Ratios $Q^{AS}_{i,1}$ of every AS spectra divided by the spectrum of group 1 (0.2 mW/µm²). Dashed black lines: Fits using Eq. 3. (d) Calculated temperature as a function of the excitation irradiance. Solid red line: mean photothermal coefficient $\langle\beta\rangle = 63$ K µm²/mW with a standard deviation of 3 K µm²/mW. Grey lines: Values obtained for ten repetitions of the measurement.

Figure 2b shows the AS spectra $I_i^{AS}$, normalized by the incident irradiance of $I_i^{exc}$. They do not overlap because they correspond to different temperatures of the Au NP. Then, ratios $Q_{i,j}^{AS} = \frac{I_i^{AS}}{I_j^{AS}}$ between every pair of spectra are computed, which including equation (2) take the form:

$$Q_{i,j}^{AS} = \frac{I_i^{AS}}{I_j^{AS}} = \frac{I_i^{exc}}{I_j^{exc}} \frac{e^{\frac{E(\lambda)-E(\lambda_{exc})}{k_B T_j}} - 1}{e^{\frac{E(\lambda)-E(\lambda_{exc})}{k_B T_i}} - 1} \qquad (3)$$

Conveniently, this expression does not depend on the intrinsic PL emission spectrum $f_{PL}$. If the range of explored temperatures is not so large, it can be assumed that the temperature of the NP increases linearly with excitation intensity.

$$T_i = T_0 + \beta\, I_i^{exc} \qquad (4)$$

where $T_0$ is the temperature of the particle in the absence of light and $\beta$ is the photothermal coefficient. Combining Eq. 3 and 4 gives

$$Q_{i,j}^{AS} = \frac{I_i^{exc}}{I_j^{exc}} \frac{e^{\frac{E(\lambda)-E(\lambda_{exc})}{k_B[T_0+\beta I_j^{exc}]}} - 1}{e^{\frac{E(\lambda)-E(\lambda_{exc})}{k_B[T_0+\beta I_i^{exc}]}} - 1} \qquad (5)$$

which is an expression with two free parameters, $T_0$ and $\beta$. If one is known, the other can be calculated from a fit to the data. As an example, Figure 2c shows all ratios $Q_{i,1}^{AS}$ using irradiance $j=1$ (the lowest) as a reference, together with fits to Eq. 5 using $T_0 = 295$ K, the room temperature during measurements. For each pair of irradiances $\{i,j\}$ a value of $\beta_{i,j}$ is extracted from a fit to $Q_{i,j}^{AS}(\lambda)$. Then, the average value $\beta$ can be determined, which enables the determination of the NP temperature as a function of the irradiance through Eq. 4. The reproducibility of the method was tested by determining $\beta$ from ten consecutive hyperspectral images of the same NP. Figure 2d shows the temperature vs. irradiance curves corresponding to each of the individual determinations of $\beta$, along with its mean value of $\langle\beta\rangle = 63$ K µm²/mW. The standard deviation of the ten measurements was of 3 K µm²/mW (4%).

In summary, the photothermal coefficient $\beta$ of single NPs can be determined from one hyperspectral image. In turn, $\beta$ delivers the NP temperature for any given irradiance. It is important to remark that no characterization of $f_{PL}$ was required, eliminating possible sources of error. For example, if $f_{PL}$ is approximated by the scattering cross-section of the NP, very high and unrealistic temperatures are obtained (in the range of thousands of degrees, see Section 1.3 on the ESI).

**Using NPs as nanothermometers**

Hyperspectral images of 80 nm Au NPs were acquired at six different temperatures of the sample, $T_{set}$, ranging from 22 to 70 °C (details on Materials and Methods). The obtained data was analyzed in the two possible ways. First, as before, $\beta$ was determined at the different temperatures by fitting $Q_{i,j}^{AS}(\lambda)$ to Eq. 5, while keeping $T_0 = T_{set}$ as a fixed parameter. The resulting curves of temperature vs. irradiance corresponding to the obtained $\langle\beta\rangle$ (10 images) are shown in Figure 3a. The slightly different slopes of the curves are in agreement with the variability observed at room temperature (Figure 2d). This is better seen in Figure 3b, where $\langle\beta\rangle$ is plotted as a function of $T_{set}$. The variability of $\langle\beta\rangle$ is contained within ± 1 standard deviation around the mean value obtained at room temperature, supporting the hypothesis of $\beta$ being independent of temperature (Eq. 3).

The second way to analyze the data is to take $\beta$ as a known parameter and extract the surrounding temperature $T_0$. In this way, if the photothermal coefficient of a NP is known, the NP can be used as a nano-thermometer. Figure 3c shows the determined $T_0$ vs. $T_{set}$ using $\langle\beta\rangle$ obtained at room temperature for this NP (62.5 K µm²/mW). The determined temperature follows faithfully the externally set temperature, though it becomes less accurate at higher temperatures. This is attributed to larger measurement errors associated with heating the sample, such as faster sample drift.

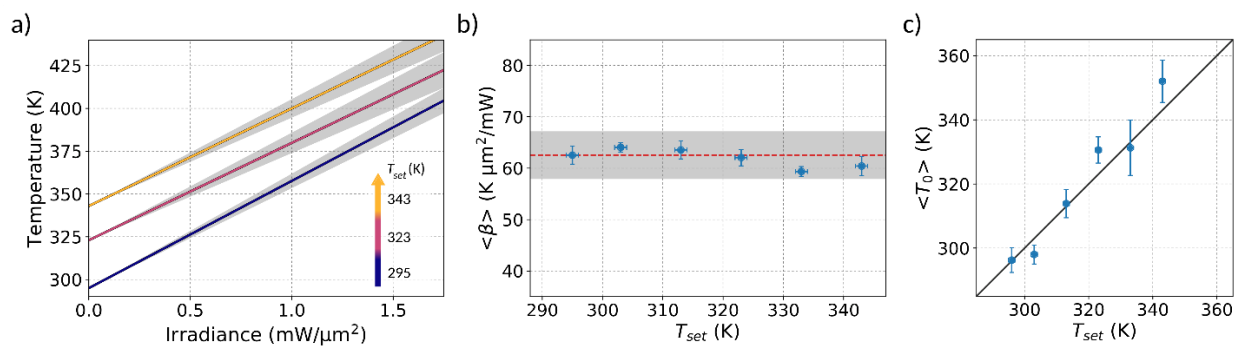

**Figure 3:** Au NPs as nanothermometers. (a) Temperature vs. irradiance for an 80 nm Au NP on a glass substrate, immersed in water at different temperatures $T_{set}$. (b) Extracted photothermal coefficient vs. $T_{set}$. The red dashed line indicates $\langle \beta \rangle$ at room temperature (62.5 K µm²/mW) and the grey band corresponds to $\pm 1$ standard deviations (4.6 K µm²/mW). (c) Determined $T_0$ vs. $T_{set}$ using a fixed photothermal coefficient $\beta = 62.5$ K µm²/mW. Error bars in (b) and (c) represent the standard error of the mean.

### Size-dependent photothermal response of spherical Au NPs

Next, the method was applied to characterize the photothermal response of spherical gold NPs of different sizes deposited onto a glass substrate and surrounded by water. NPs with four different nominal diameters of 48, 64, 80 and 103 nm were used. Arrays of individual Au NPs were fabricated through optical printing[21,48–50] (further information can be found in our previous works and in Materials and Methods). The NPs of 48, 80 and 103 nm were ultra-smooth[51–53] whereas the 64 nm ones were conventional, commercially available NPs.

An illustrative dark-field image of the optically printed Au NPs is shown in Figure 4a. Representative scattering spectra of individual NPs of each size are shown in Figure 4b (solid lines), along with the corresponding calculated spectra using Mie theory, considering the NPs immersed in water (no substrate).

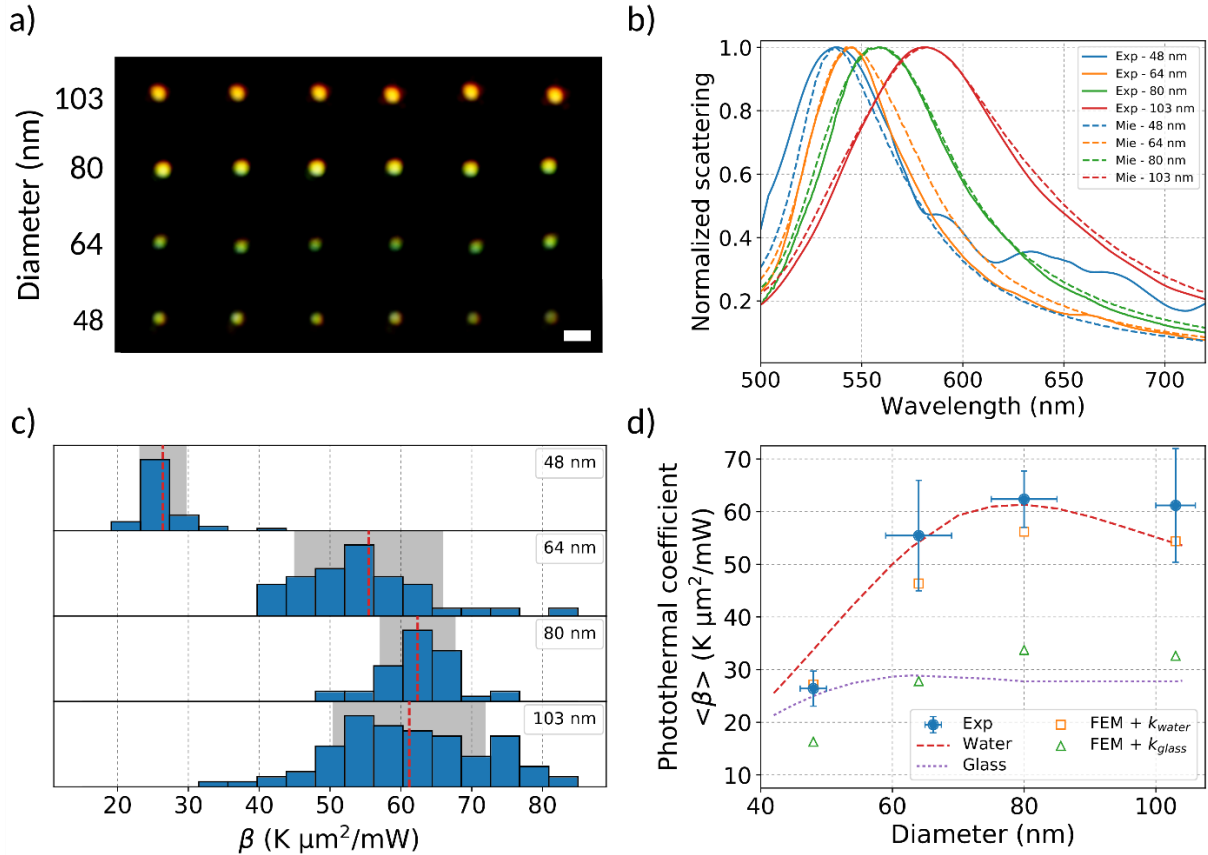

**Figure 4:** Photothermal response of spherical Au NPs as a function of size. (a) Dark-field image of spherical gold NPs optically printed on glass, immersed in water. Scale bar = 1 μm. (b) Scattering spectra of single NPs of each size. Solid lines: Experimental data. Dashed lines: Mie theory considering the NPs in water (no substrate). (c) Histograms of photothermal coefficients. The mean (red dashed lines) and an interval corresponding to ±1 standard deviations (grey band) are shown. (d) Experimental and calculated photothermal coefficient vs. diameter. Error bars in diameter represent the standard deviation of NPs size as measured by TEM and the standard deviation of $\beta$.

The photothermal coefficient $\beta$ was determined at least 40 times for each NP size, with a maximum of 90 times. Figure 4c shows the resulting histograms of $\beta$. The width of the distributions is significantly larger than the variability in the determination of $\beta$ for a single NP, which reflects the size distribution of the NPs. The narrowest distributions of $\beta$ are obtained for 48 nm and 80 nm ultra-smooth NPs, consistent with their narrower size distribution, shown in Figure S5. Figure 4d shows the mean $\langle \beta \rangle$ vs. the diameter of the NPs.

Further insight into the heat dissipation is obtained if $\beta$ is decomposed into its optical and thermal contributions:

$$\beta(\lambda) = \frac{\sigma_{abs}(\lambda)}{\gamma} \tag{6}$$

where $\sigma_{abs}$ is the absorption cross-section and $\gamma$ the effective heat dissipation factor, which depends on the geometry and thermal properties of the system. The simplest way to model $\gamma$ is to consider the NPs as a sphere with radius $a$, immersed in a homogeneous environment with thermal conductivity $\kappa$:[3]

$$\gamma = 4\pi a \kappa \tag{7}$$

The same holds for the optical properties of the NPs. The simplest way of estimating $\sigma_{abs}$ is through Mie theory, considering the NPs as gold spheres immersed in a uniform medium (no substrate). This approximate calculation reproduces well the scattering cross-section (Figure 4b). Dashed and dotted lines in Figure 4d show the predicted values of $\beta$ for the spherical NPs in homogenous environments, water or glass, using Mie theory for $\sigma_{abs}$ and Eq. 7 for $\gamma$. Remarkably, this simple calculation for water describes the experimental trend, especially for the larger NPs. This appears reasonable since for large NPs most of the surface is in contact with the water. A more accurate calculation of the absorption cross-section, considering the presence of the substrate, was performed using a finite element method solver (FEM, see Section 2.2 of the ESI for further details). Figure 4d shows the predicted values of $\beta$ using FEM absorption cross-sections and $\gamma$ as Eq. 7, in water (orange squares) and glass (green triangles). Again, a reasonable agreement between calculations and experimental results is obtained for $\gamma$ in water. However, the predicted values are smaller than the experimental ones for large Au NPs, indicating that the simple model of Eq. 7 overestimates $\gamma$. This result is rather surprising, as it is the opposite of expected. This model ignores the glass substrate which is a better thermal conductor than water ($\kappa_{\text{glass}} = 1$ W/K m, $\kappa_{\text{water}} = 0.6$ W/K m). Thus, any correction made to include the substrate would lead to a larger $\gamma$, and as a consequence to a smaller $\beta$, enlarging, even more, the differences with the experimental data. The only way to obtain a lower value for $\gamma$ would be to consider the thermal resistances of the Au-water, Au-substrate and water-substrate interfaces (Kapizka resistances $R_K$). However, considering these three parameters require precise knowledge of the interfaces and the geometric boundaries, and thus increase the complexity of the model. To summarize, the results presented so far emphasize the importance of interfaces in heat transport at the nanoscale, highlighting the necessity of *in situ* experimental methods.

**Photothermal response of 80 nm Au NPs on different substrates**

To gain further insight into the role of substrates in heat dissipation, the photothermal response of 80 nm Au NPs was characterized on two highly conductive substrates: sapphire and graphene-coated glass. Sapphire has a thermal conductivity an order of magnitude higher than glass ($\kappa_{\text{sapphire}} \cong 20$ W/K m). Graphene is known for its outstanding thermal conductivity. When a graphene monolayer is deposited onto SiO$_2$, its in-plane thermal conductivity can be up to two orders of magnitude higher than glass. For example, Seol *et al.*[54] reported $\kappa_{\text{graphene@SiO2}} = 600$ W/K m and Li *et al.*[55] reported 840 W/K m.

Arrays of 80 nm ultra-smooth Au NPs were optically printed onto these two substrates (see Materials and Methods for details). Sapphire substrates were treated with polyelectrolytes following the same procedure used with the glass substrates. A monolayer of graphene was deposited onto glass substrates using a wet transfer method (further details in Materials and Methods). Polyelectrolyte functionalization was not necessary in the graphene substrates because nonspecific deposition of Au NPs was not observed during the optical printing process, probably due to hydrophobic repulsion. In the graphene substrates, prior to the optical printing of NPs, Raman spectra were acquired to confirm the presence of a monolayer of graphene in the printing area (Figure S7a). The characteristic Raman peaks of graphene are still visible in the PL measurements used for thermometry (Figure S7b), indicating that graphene is not damaged by the optical printing process.

Figure 5a shows example dark-field images of the arrays of NPs on the different substrates. Figure 5b shows the mean temperature increase vs. irradiance whereas Figure 5c shows the histograms of photothermal coefficients for the NPs on each substrate.

The mean photothermal coefficient for the NPs supported on sapphire was $(51 \pm 1)$ K μm²/mW. This value is lower than on glass, which is reasonable given its higher thermal conductivity. Interestingly, a substrate with thermal conductivity 20 times higher leads to a reduction of $\langle \beta \rangle$ of only 18%. Sapphire has a significantly higher refractive index than glass, and therefore the absorption spectrum of the Au NPs is red-shifted with respect to glass. FEM simulations were performed to take this effect into account. $\sigma_{abs}$ (at $\lambda = 532$ nm) was estimated to be $16.9\ 10^3$ nm² on glass and $18.9\ 10^3$ nm² on sapphire. Considering this and using Eq. 6, the calculated heat dissipation factor $\gamma$ for a NP on sapphire is 36% higher than for glass. These results are consistent with the ones obtained for 80 nm NPs on glass, where it was observed that the

**16**

thermal transport is dominated by the surrounding water. It is important to note that the glass and sapphire surfaces were treated in the same manner with polyelectrolytes. Thus, it is reasonable to expect a similar surface chemistry and Kapitza resistance $R_K$[56,57] between the substrate and the water. On the contrary, $R_K$ between the NP and the substrate might be different.

The obtained photothermal coefficient for the NPs supported on graphene-coated glass is $\langle \beta \rangle = (77 \pm 1)\ \text{K}\,\mu\text{m}^2/\text{mW}$. Surprisingly, for a given irradiance level, the NPs on graphene reach temperatures 24% higher than on bare glass. It was observed that the scattering spectra of NPs on the graphene substrate red-shifts with respect to glass, on average by 12 nm (Figure S8). Using Mie theory with an effective refractive index for the medium, it was estimated that $\sigma_{abs}(\lambda = 532\ nm)$ is reduced by 7% with respect to glass (Figure S8b). Taking this into account, it was estimated that the heat dissipation factor $\gamma$ for a NP on graphene is 75% of the value for bare glass. Instead of improving heat dissipation, the graphene monolayer acts as an insulator, in spite of its great heat conductivity. This is ascribed to the $R_K$ at the graphene interface. Contrary to the hydrophilic surface of polyelectrolyte-coated glass, graphene-coated glass is highly hydrophobic.[58] $R_K$ is inversely proportional to the work of adhesion[56,59], and therefore a higher $R_K$ resistance is expected between the water and the hydrophobic graphene-coated glass. Moreover, very high values of $R_K$ between graphene and $SiO_2$ interfaces have been measured,[60] including a dramatically five-order of magnitude increase in $R_K$ when the graphene is not in full contact with glass but presents some corrugation of nm-sized height.

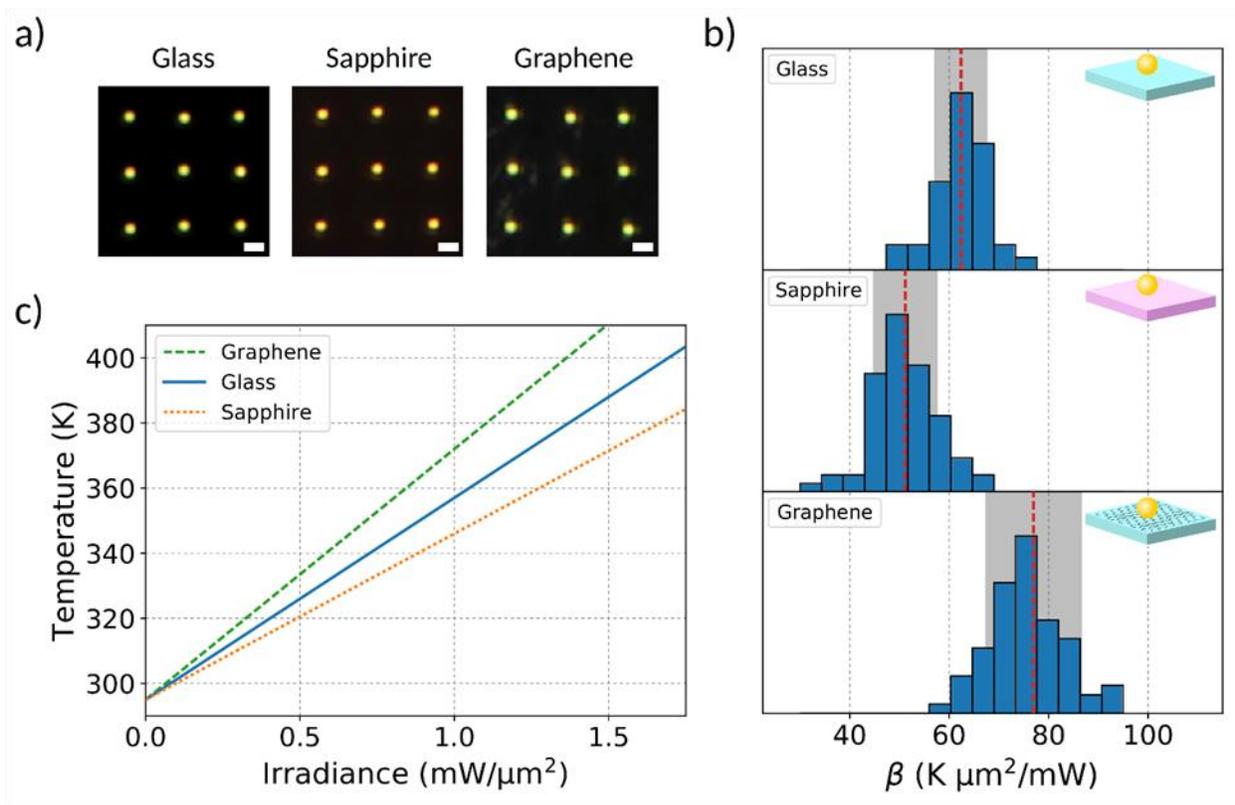

**Figure 5:** The role of the substrate in the photothermal coefficient. (a) Example dark-field images of the optically printed arrays of NPs on the different substrates. (b) Temperature vs. irradiance for 80 nm Au NP on glass, sapphire and graphene substrates, immersed in water. (c) Histograms of photothermal coefficients. The mean photothermal coefficients (red line) and corresponding standard error of the mean are $\langle \beta \rangle_{\text{glass}} = (62 \pm 1)$, $\langle \beta \rangle_{\text{sapphire}} = (51 \pm 1)$ and $\langle \beta \rangle_{\text{graphene}} = (77 \pm 1)$ K µm²/mW. The grey band corresponds to a $\pm 1$ standard deviations interval.

CONCLUSIONS

A new implementation of anti-Stokes nanothermometry was presented that allows the measurement of the photothermal coefficient $\beta$ of single NPs from an individual hyperspectral photoluminescence image. Unlike previous implementations, this thermometry method can be used to probe *in situ* the photothermal response of any type of NP, provided it delivers detectable anti-Stokes emission. It uses excitation at a single wavelength and fixed power, and does not require any extra characterization or prior knowledge about the NP scattering or PL spectra. The method retrieves $\beta$ with a precision of 4%. Alternatively, if $\beta$ is known, or previously determined, the same method serves to use NPs as nano-thermometers to sense the temperature of the surroundings.

Using this method, the photothermal response of spherical Au NPs on glass substrates, immersed in water was characterized. It was found that for NPs with diameters larger than 50 nm, $\beta$ can be modeled by a homogeneous environment of just water, indicating only a minor role of the substrate in heat dissipation. This finding was consistent with measurements on 80 nm Au NPs deposited on substrates with nominally higher thermal conductance. On sapphire, $\beta$ was found to be 18% smaller than on glass. Taking into account the change in absorption induced by the sapphire substrate, it was estimated that the heat dissipation for the NPs on sapphire was only 36% higher than for NPs on glass, despite sapphire has a thermal conductivity 20 times higher. NPs on graphene-coated glass presented a $\beta$ 24% larger than on bare glass. Considering the change in absorption induced by the graphene substrate, it was determined that NPs on graphene-coated glass dissipate heat 25% less efficiently than on bare glass. This is a rather surprising result, at least at first sight: including a material with extremely high heat conductivity leads to a NP more thermally isolated. However, this can be explained by the presence of high thermal resistances at the hydrophobic graphene interface. Further studies to disentangle these parameters would unlock the potential of graphene materials for thermal management in the nanoscale.

Overall, the results of this work reveal the complexity of heat dissipation around NPs and highlights the value of *in situ* experimental methods able to quantitatively assess the photothermal response of nanosystems. At the same time, they provide a guideline and a method for future research. In this context, we believe that the *in situ* thermometry method presented here will be useful to address open questions about the role of temperature in plasmon-assisted applications, such as photocatalysis or optical manipulation of NPs.

MATERIALS AND METHODS

Ultra-pure water was employed in all cases (18 MΩ·cm, Milli-Q®, Millipore).

**Gold nanoparticles:** Different sized AuNPs capped with CTAB (positively charged) were prepared and purified accordingly to Ref [52]. The diameters were (48 ± 2) nm, (80 ± 5) nm and (103 ± 3) nm). Au NPs of (64 ± 5) nm in diameter stabilized with NPC privative capping agent (negatively charged) were purchased from NanoPartz. All NPs sizes were characterized by transmission and/or scanning electron microscopy (CMA, FCEyN, University of Buenos Aires, Argentina). In order to reach the adequate concentration to optically print the NPs[21,48,49], the positively charged NPs were diluted using a 2 mM CTAB solution, and a 1.5 mM NaCl solution was employed for the 64 nm NPs.

**Substrate functionalization:** Polydiallyldimethylammonium chloride (PDDA, MW ∼ 400.000 – 500.000) and sodium polystyrene sulfonate (PSS, MW = 70.000) were purchased from Sigma-Aldrich and used without further purification. Soda-lime glass (Paul Marienfeld GmbH & Co, DE) and sapphire (PI-KEM, UK, Aluminum Oxide <0001>) substrates were cleaned using Hellmanex III® 0.2% for 10 min in an ultrasound bath at room-temperature. After rinsing with deionized water and acetone, they were dried in an oven at 85 °C for 2 hs. Their surface was plasma-activated using a plasma cleaner (Diener, Zepto) at 75 W and 0.5 mbar for 3 min. Finally, positive-charged substrates were produced by immersion of the glass/ sapphire in a PDDA solution (1 mg/ml in 0.5 M NaCl) for 15 min and afterwards they were rinsed with ultrapure water several times. Substrates were stored in water for no longer than one week. Positively charged substrates were used to print

**20**

the CTAB capped Au NPs. For the negatively charged Au NPs (64 nm), the previously functionalized PDDA substrates were immersed in a PSS solution (1mg/ml in 0.5 M NaCl) for 15 minutes and rinsed again with ultrapure water.

**Graphene transfer on glass substrates:** Monolayer graphene sheets were deposited on soda-lime glasses by a wet transfer method.[61] First, glass substrates were cleaned as previously described. Graphene-on-copper foils by chemical vapor deposition (CVD) were obtained from Graphenea Inc[62]. Graphene-copper-foils were coated with polystyrene (PS, MW = 290.000 g/mol) by drop casting from a solution of PS in toluene and dried at 75 °C. Then, the underlying copper support was etched using a solution of hydrochloric acid (1.4 M) and hydrogen peroxide (0.5 M). PS-graphene sheets were transferred onto the glass substrate and dried at 75 °C. Finally, PS was removed using toluene and the graphene-coated substrates were annealed at 200 °C overnight.

**Experimental methods:** Au NPs were optically printed onto each substrate according to the process described in our previous works.[21,48,49] Briefly, the Au NP suspension is placed on top of the functionalized substrate having the same charge as the NPs in an open chamber and printed using a 532 nm laser (Ventus, Laser Quantum) using a 60x water-immersion (Olympus) objective with a NA 1 (beam waist $\omega_0 = (342 \pm 5)$ nm). Printing power was adjusted to each system in order to avoid morphological changes during the optical printing process. Typically, 0.9 mW was used to print onto glass, 0.78 mW for sapphire and 1 mW for graphene. For nanothermometry, laser power was kept rather low to elude long-term irradiation damage. Hyperspectral confocal images of 48 nm, 64 nm, 80 nm and 103 nm AuNPs were acquired using 0.80 mW, 0.36 mW, 0.31 mW and 0.20 mW, respectively. The heating stage was built using two heaters (MP800, Caddock Electronics) in contact with a stainless-steel sample holder. A temperature sensor (LM35, Texas Instruments) was used to monitor the stage temperature. A PID controller was implemented using an Arduino Uno R3 board for data acquisition and control via a Python-based software. A settling time of 30 min was left after the temperature set-point was reached and before the acquisition of hyperspectral images at different temperatures. The entire process of optical printing, hyperspectral imaging and analysis was fully automated by Python routines.

ASSOCIATED CONTENT

**Supporting Information**

The Electronic Supporting Information (ESI) contains details about linearity of Stokes emission, comparison of different temperature-dependent distribution functions and their ratios, estimation of the difference between lattice and electronic temperatures, histograms of maximum scattering wavelength of different sized Au NPs, information about calculations of optical properties and Raman spectra of graphene. The following files are available free of charge (pdf).

AUTHOR INFORMATION

**Corresponding Author**

*J.Gargiulo@physik.uni-muenchen.de

*Emiliano.Cortes@lmu.de

*fernando.stefani@cibion.conicet.gov.ar

**Present Addresses**

† Instituto de Nanosistemas, UNSAM-CONICET, Av. 25 de Mayo 1021, San Martín 1650, Argentina.

**Author Contributions**

The manuscript was written through contributions of all authors. All authors have given approval to the final version of the manuscript.

ACKNOWLEDGMENT

M.B., I.L.V., L.P.M and F.D.S are thankful for the financial support from ANPCyT (PICT 2013-0792, PICT 2014-3729 and PICT 2017-0870). M.B. acknowledges CONICET for a postdoctoral scholarship and L.P.M. for a doctoral scholarship. J.G. acknowledges the European Commission for the Marie-Sklodowska-Curie action (797044) and the PRIME programme of the German Academic Exchange Service (DAAD) with funds from the German Federal Ministry of Education and Research (BMBF). D.P. and V.G. thank Prof. Dr. Kannan Balasubramanian for his help in the graphene transfer method. S.S. acknowledges funding from the German Cancer Foundation (DKH, Deutsche Krebshilfe) within the Priority Program 'Visionary Novel Concepts in Cancer Research'. S.A.M and E.C acknowledge funding and support from the Deutsche Forschungsgemeinschaft (DFG, German Research Foundation) under Germany´s Excellence Strategy – EXC 2089/1 – 390776260, the Bavarian program Solar Energies Go Hybrid (SolTech) and the Center for NanoScience (CeNS). E. C. acknowledges support from the European Commission through the ERC Starting Grant CATALIGHT (802989).

TABLE OF CONTENTS

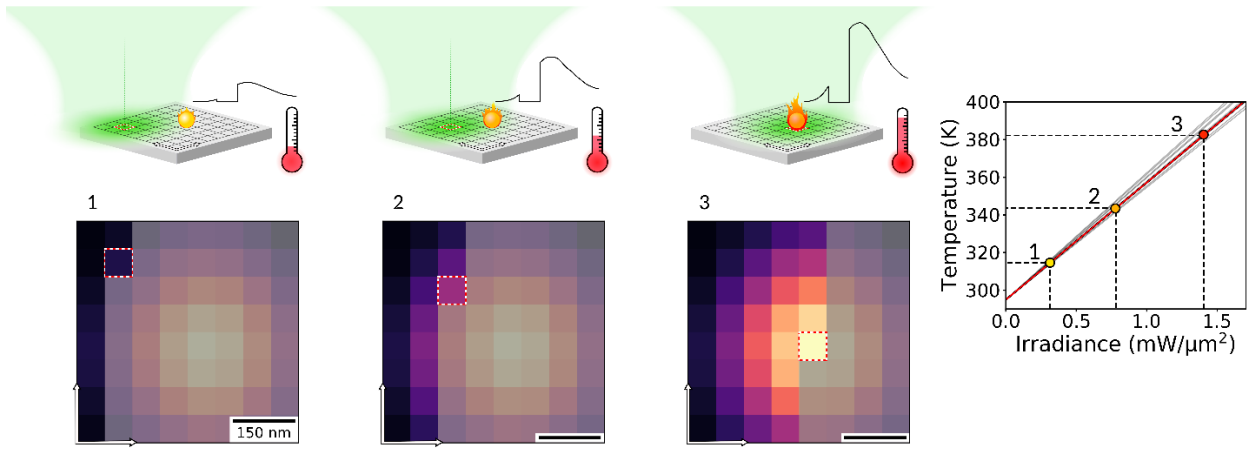